\documentstyle[epsfig]{mn}

\ifCUPmtlplainloaded \else

\fi

\begin{document}

\title{Radio flares and plasmon size in Cygnus X-3}

\author[Ogley et al.]{
R. N. Ogley,$^{1,2}$\thanks{E-mail: rno@astro.keele.ac.uk}
S. J. Bell Burnell$^{1}$,
R. E. Spencer$^{3}$,
S. J. Newell$^{3}$,
\newauthor
A. M. Stirling$^{4}$ and
R. P. Fender$^{5}$\\
$^{1}$ Department of Physics and Astronomy, The Open University, Walton
Hall, Milton Keynes, Buckinghamshire, MK6 7AA, UK\\
$^{2}$ Present address: Department of Physics, Keele University, Staffordshire, ST5 5BG, UK\\
$^{3}$ University of Manchester, Jodrell Bank Observatory, Macclesfield, Cheshire, SK11 9DL, UK\\
$^{4}$ CFA, University of Central Lancashire, Preston, PR1 2HE, UK\\
$^{5}$ Astronomical Institute `Anton Pannekoek' and Center for
High-Energy Astrophysics, University of Amsterdam,\\
Kruislaan 403, 1098 SJ Amsterdam, the Netherlands}

\maketitle

\begin{abstract}
We have observed a number of minor radio flares in Cyg X-3 using the
MERLIN array.  Photometric observations show the system to be highly
active with multiple flares on hourly timescales over the one month
observing programme.  Analysis of the source's power spectrum show no
persistent periodicities in these data, and no evidence of the 4.8 hr
orbital period.  An upper limit of 15 mJy can be placed on the
amplitude of any sinusoidal varaition of source flux at the orbital
period.  The brightness temperature of a flare is typically $T_{\rm b}
\geq 10^{9}$-$10^{10}$ K, with a number of small flares of 5 minute
duration having brightness temperatures of $T_{\rm b} \geq {\rm few}
\times 10^{11}$ K.  For such a change in flux to occur within a
typical 10 minute timescale, the radiation must originate from
plasmons with a size $\leq$ 1.22 AU.  This emission is unlikely to
originate close to the centre of the system as both the jets and
compact object are buried deep within an optically thick stellar wind.
Assuming a spherically symmetric wind, plasmons would become visible
at distances $\sim$ 13 AU from the core.
\end{abstract}


\section{Introduction}

The energetic X-ray binary Cygnus X-3 has been studied in great detail
since its discovery in 1967 (Giacconi et al.\ 1967).  It is an active
source with measurements in the radio, sub-mm, infrared, X-ray and
gamma-rays all of which indicate a highly variable source.

Following infrared spectroscopic measurements (van Kerkwijk et
al. 1992; Fender, Hanson \& Pooley 1999), the accepted morphology of
Cyg X-3 is a compact object in orbit around a Wolf-Rayet star of the
WN4-5 type.  A measured infrared and X-ray orbital period for the
binary of 4.8 hr indicates a very compact system $(\leq 10\;{\rm
R_{\odot}})$ within which a large degree of interaction takes place.
The proximity of a compact object to a star with a strong wind
provides both accretion onto the compact object, but also a medium of
varying optical depth through which jets, formed from accreted
material, must travel.  Any emission from the jets cannot be observed
directly until the wind becomes optically thin to the radiation in the
jets.  This creates a time-lag between radiation at different
frequencies in the jet (typically radio and sub-mm emission) and
between jet and central infrared and X-ray emission.

One can use the different radiation regimes to investigate different
properties of the system.  For example, the low-frequency radiation at
radio and sub-mm frequencies comes from synchrotron emitting electrons
in the jets, and this has been used by Ogley et al.\ (in prep.) to obtain
an upper-limit to the magnetic field in the jets of 260 $\mu$T (2.60
G) at 100 $\rm R_{\sun}$ from the core.  As one observes higher
frequencies, the emission changes from synchrotron to thermal
free-free, which now comes from the wind in the system and not the
jets.  An example of the wind emission observed by the {\it ISO}
satellite is given in Ogley, Bell Burnell \& Fender (2001), and a
complete spectrum showing the different emission mechanisms from radio
to infrared is given in Ogley et al.\ (in prep.) Fig.\ 6.  However, a
thorough investigation using a single radiation regime can also
provide a great deal of information.

The accretion process is not constant, but is probably variable in
both density and mass-transfer rate.  This causes a change in the
number of electrons which are available for acceleration in the jet,
which alters the synchrotron luminosity.  Radio photometry monitors
the change in jet emission following a change in the Wolf--Rayet state
and three distinct states of emission in the unresolved core of the
system have been identified (Waltman et al.\ 1995):
\begin{enumerate}
\item{For the majority of the time the system is in a quiescent radio
state with GHz flux densities around 100 mJy.}
\item{Once or twice a year the system undergoes a major flare.  The
radio flux density increases by 10-100 times on a timescale of around
a day.  A major flare is very energetic and lasts for days or weeks,
exponentially decaying back to quiescence.  Identified immediately
before a major flare is a period of quenching during which the flux
density drops to around 10 mJy and remains at this lower value for up
to a week.}
\item{An intermediate state exists between the high flux of a major
flare ( $>$ 1 Jy) and the normal quiescent flux of 100 mJy.  This is
the state of minor flaring and can last for several weeks.  During a
minor flare state, the system can undergo multiple short-duration
flares and is highly active.}
\end{enumerate}

Continuous monitoring determines the radio state of Cyg X-3, and
allows detailed investigation of the jet mechanism and structure
during these different states.

\section{The radio observations}

Daily radio monitoring was carried out using the Ryle Telescope in
Cambridge and the Green Bank Interferometer (GBI) in West Virginia.
These observatories provided photometry at 2.3 and 8.3 GHz (GBI) and
15 GHz (Ryle).  Identification of a minor flare in 1996 November
triggered a set of interferometric observations using the MERLIN
array.  The MERLIN observations were scheduled at C-band (5 GHz) with
6 epochs of 12 hours in duration each separated by one week.  The aim
of these observations was to map the emission from a plasmon as it
travelled from the centre of the binary, as observed by Newell (1995).
At a frequency of 5 GHz a plasmon at 10 kpc, travelling at a speed of
0.35 $c$ in the plane of the sky, would separate from the core at one
MERLIN beam-width per epoch $(\simeq 50\;{\rm mas})$.  This jet
velocity, taken from the apparent transverse motion measured by Spencer
et al.\ (1986) represents a lower-limit to the separation per epoch.
Other measurements have indicated higher velocities of components
close to the plane of the sky ($\simeq 0.6c$ from Mart\'\i~et al.\
(2000) and $\simeq 0.8c$ from Mioduszewski et al.\ (2000)).  These
larger apparent velocities would therefore yield greater separations
per MERLIN epoch.

\begin{figure}
\centering
\epsfig{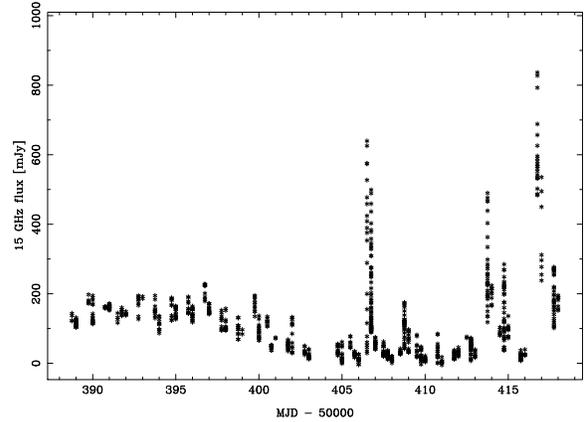}
\caption{15 GHz radio photometry from the Ryle telescope which triggered the
start of our MERLIN observation programme.  A minor flare occurred on
MJD 50407 and our MERLIN observations were triggered from MJD 50419
until MJD 50459 as discussed in the text.  See also Fig.\ 2.}
\label{ryle-phot}
\end{figure}

Fig.\ \ref{ryle-phot} shows the radio photometry from the Ryle
telescope at 15 GHz which was used as a trigger for the MERLIN
observations.  Cyg X-3 was in quiescence until MJD\footnote{In this
paper we use the definition of modified Julian date to be MJD = JD $-$
2\,400\,000.5, so integer values of MJD fall on 0000 UT.} 50400 when it
underwent a mild quenching period (flux around 50 mJy).  Shortly after
the quenching, a minor flare occurred on MJD 50407 which signalled the
start of a minor flare period.  Subsequent minor flares occurred at
random times after this date.

Observations were taken using the standard 5 GHz MERLIN continuum
setup.  A 15 MHz bandwidth was used for observations of the target
source Cygnus X-3, a phase calibrator source 2005+403, a flux
calibrator 3C84, and a point source calibrator, either OQ 208 or
0552+398.  To calibrate the flux easily, observation of a point source
of known flux is required.  However, on the longer MERLIN baselines
all the point sources are variable in flux, and all the constant flux
sources are resolved.  One has to observe a point source over all
baselines and calibrate the flux scale by measurements of a source of
known flux using the short baselines only.  Details of this
calibration procedure are given in the MERLIN user's guide, Thomasson
et al.\ (1993) and Ogley (1998).

\begin{figure}
\epsfig{file=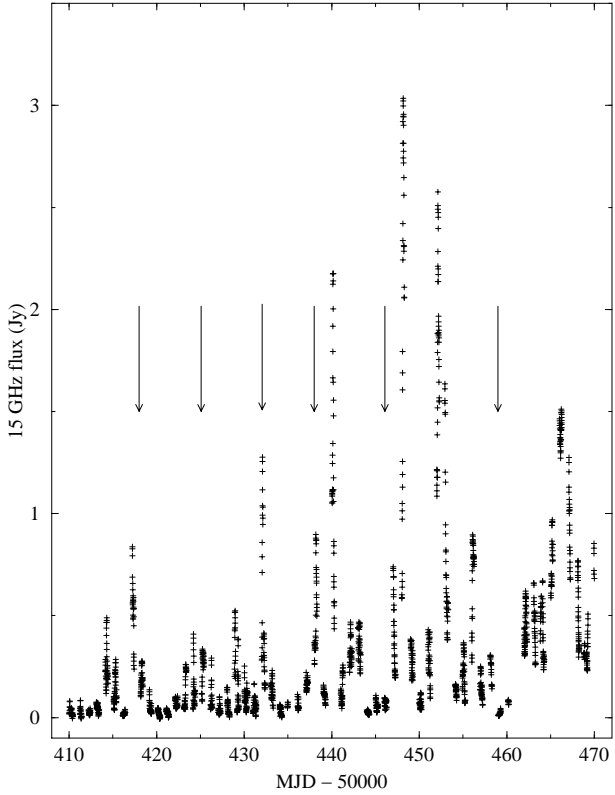, width=3.2in}
\caption{Times of the MERLIN observations indicated by arrows
overlaying a plot of Ryle photometry at 15 GHz.}
\label{Merlin_epochs}
\end{figure}

The source and observational details for all the epochs are given in
Table \ref{obs_characteristics}.
\begin{table}
\caption{Date and calibration sources for the MERLIN observations.}
\begin{tabular}{llll}
\hline
	&			& \multicolumn{2}{c}{Calibrators} \\
MJD	& Date			& Point source, flux	& Phase \\
\hline
50419	& 01 December 1996	& OQ 208	& 2005+403	\\
50425	& 07 December 1996	& OQ 208	& 2005+403	\\
50432	& 15 December 1996	& 0552+398	& 2005+403	\\
50438	& 21 December 1996	& OQ 208	& 2005+403	\\
50446	& 28 December 1996	& OQ 208	& 2005+403	\\
50459	& 11 January 1997	& 0552+398	& 2005+403	\\
\hline
\end{tabular}
\label{obs_characteristics}
\end{table}
The times of the MERLIN observations are shown in Fig.\
\ref{Merlin_epochs} together with Ryle photometry at 15 GHz around the
same time.  The MERLIN observations are indicated by the arrows.

\section{MERLIN photometry}

\begin{figure}
\epsfig{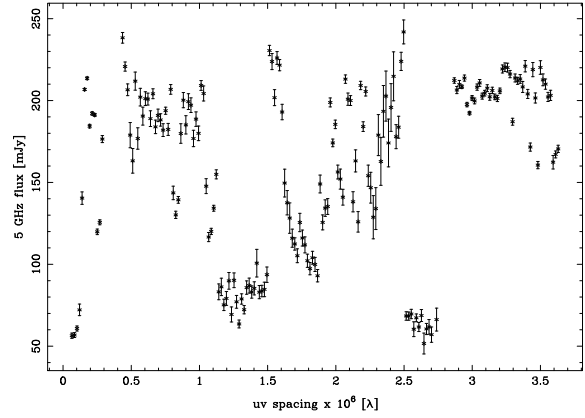}
\caption{A plot of flux against $uv$ distance for the entire first
epoch.  There is significant, and unstructured, variation of flux
which is unlikely to be caused by source structure and is due to
changes in the source flux over the epoch.}
\label{uv_variations}
\end{figure}

Because of the variable nature of the source at the time of the MERLIN
observations, a robust map of the source is almost impossible to
obtain.  Variations in observed amplitude can be due to two things: a
variable core; or emission away from the phase centre.  To reconstruct
an image, the assumption that any variation in the fringe amplitude
with hour angle is due to the interferometer response to the source
structure only is made.  However, this is violated if the source
varies within the time needed for aperture synthesis.  For our
observations, a direct reconstruction of the source results in many
artifacts including jet-like emission from a central core.  The
variability of the source is shown in Fig.\ \ref{uv_variations} which
plots the measured flux against baseline length or $uv$ spacing.  One
can see that the source is highly variable over all baseline distances
and no accurate reconstruction of the source brightness distribution
can be obtained.

The plot shown in Fig.\ \ref{uv_variations} does allow some quantative
evaluation.  A resolved source would show a decrease in measured flux
with increasing $uv$ spacing, whereas during a flare the measured flux
is approximately constant over all $uv$ separations.  In Fig.\ 3 the
measured flux is high at the longest baselines suggesting that the
source is unresolved.  A source that is unresolved at a $uv$ spacing
of 3.5 M$\lambda$ at 5 GHz has an angular size of less than 70 mas.
This is consistent with a scatter-broadened size of Cygnus X-3, which
has a strong dependence on frequency (Wilkinson, Narayan \& Spencer
1994) and has been measured at 5 GHz to be $\simeq 20-30$ mas
(Mioduszewski et al.\ 2000).

\begin{figure}
\epsfig{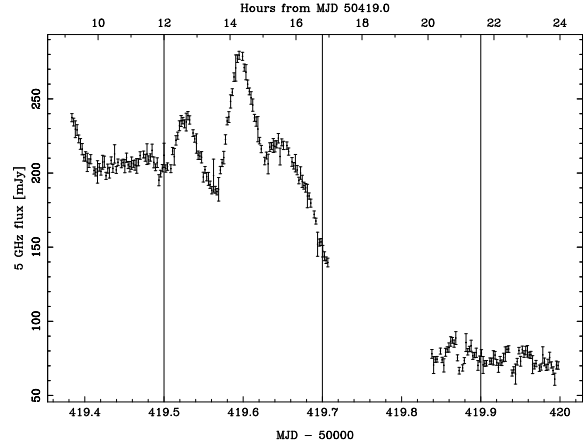}
\caption{Cyg X-3 photometry on 01 December 1996.  Time bins are 3.4
min.  Vertical lines indicate time of X-ray minimum from the
ephermeris by Matz et al.\ (1996).}
\label{photometry_1}
\end{figure}

\begin{figure}
\epsfig{file=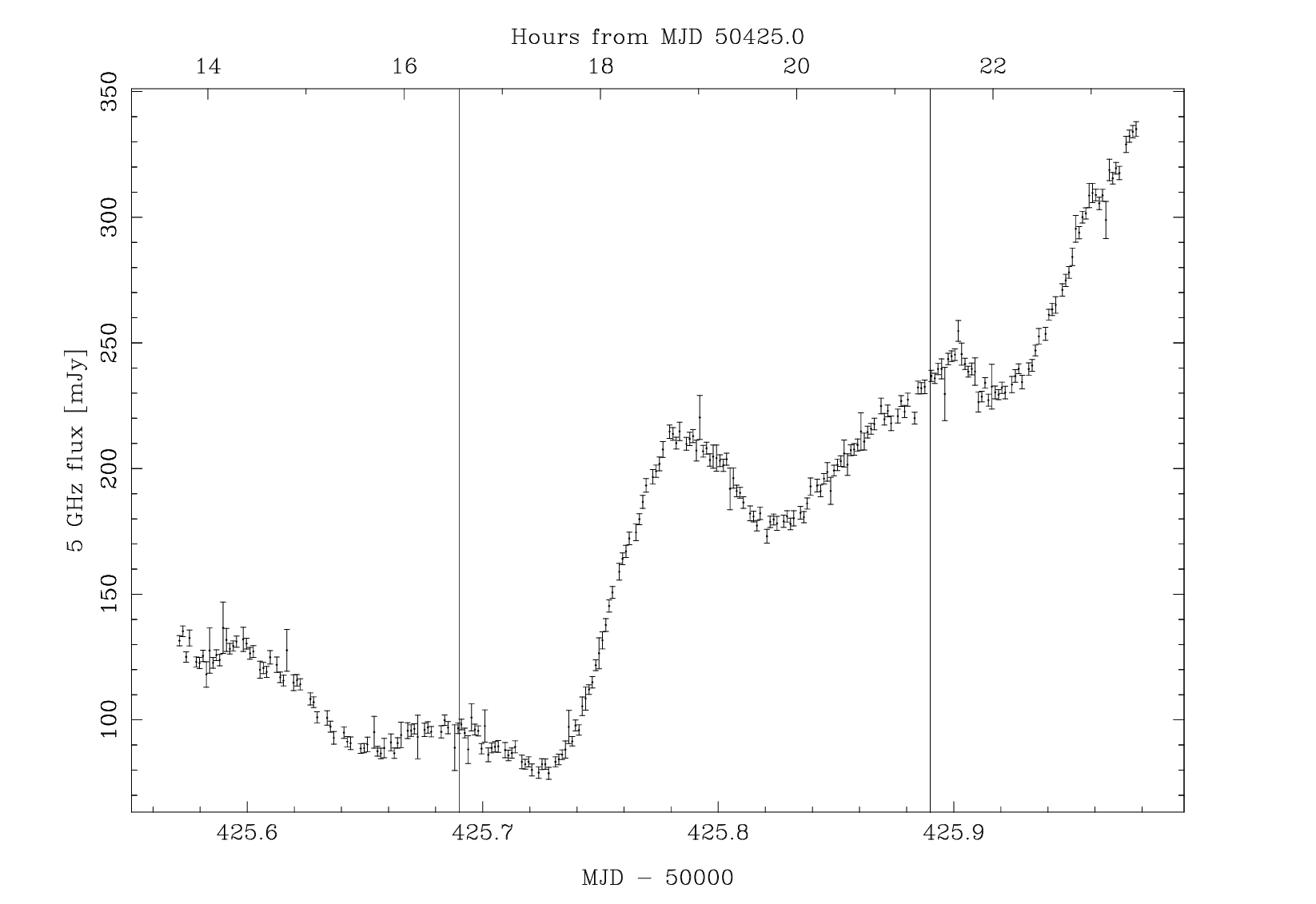, angle=-90, width=3.5in}
\caption{Cyg X-3 photometry on 07 December 1996.  Time bins are 2.2 min.}
\label{photometry_2}
\end{figure}

\begin{figure}
\epsfig{file=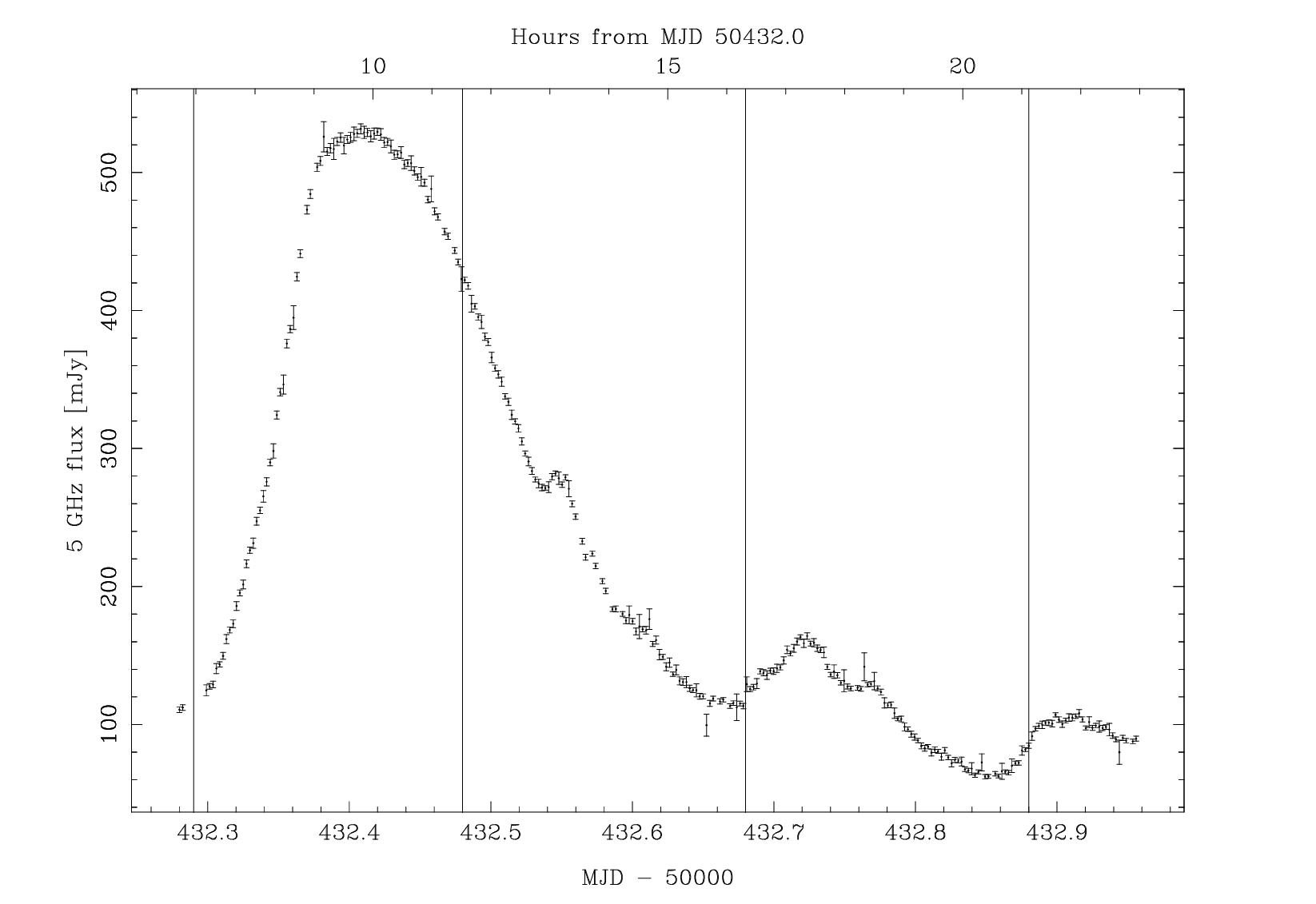, angle=-90, width=3.5in}
\caption{Cyg X-3 photometry on 15 December 1996.  Time bins are 3.8 min.}
\label{photometry_3}
\end{figure}

\begin{figure}
\epsfig{file=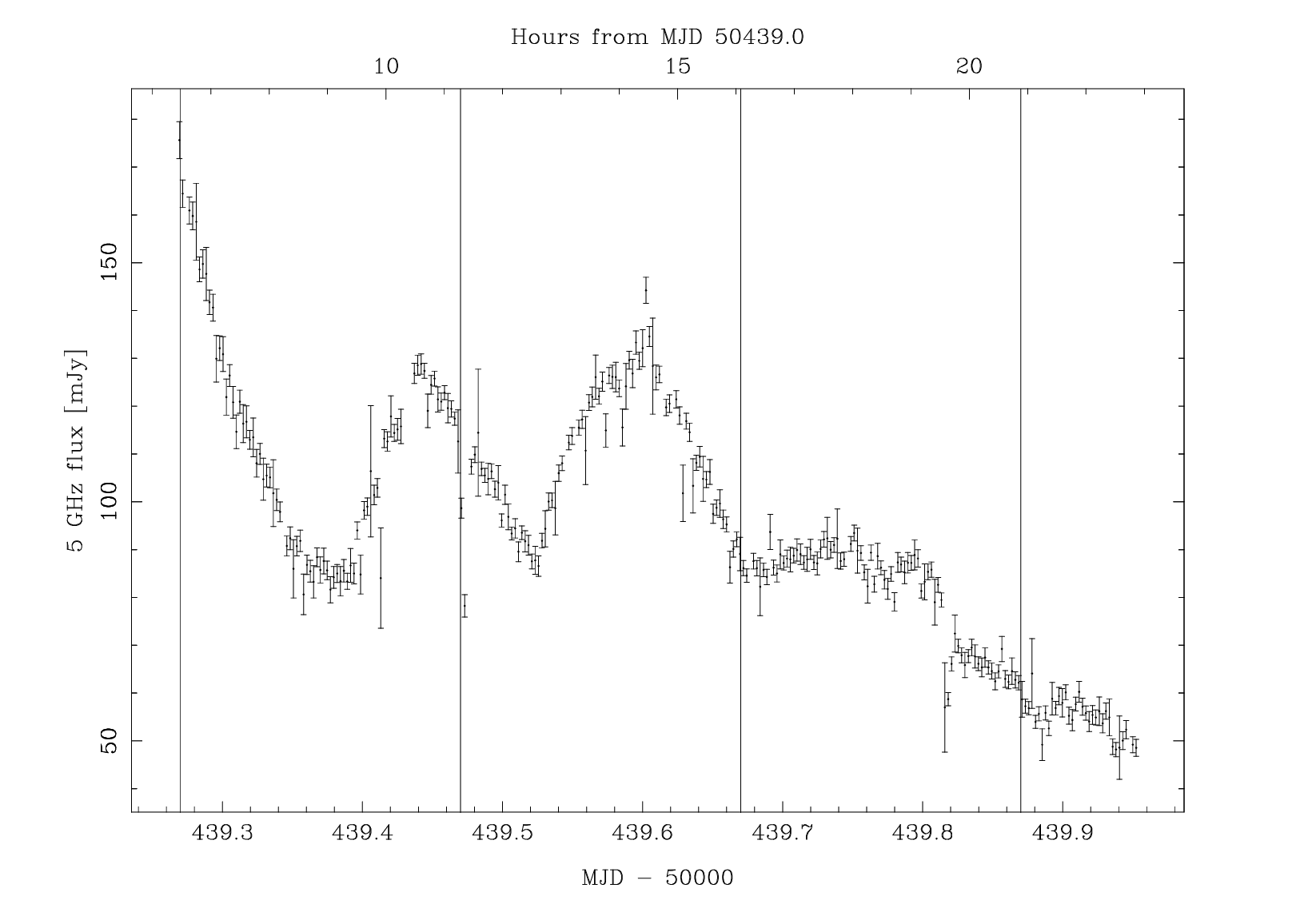, angle=-90, width=3.5in}
\caption{Cyg X-3 photometry on 21 December 1996.  Time bins are 4.0 min.}
\label{photometry_4}
\end{figure}

\begin{figure}
\epsfig{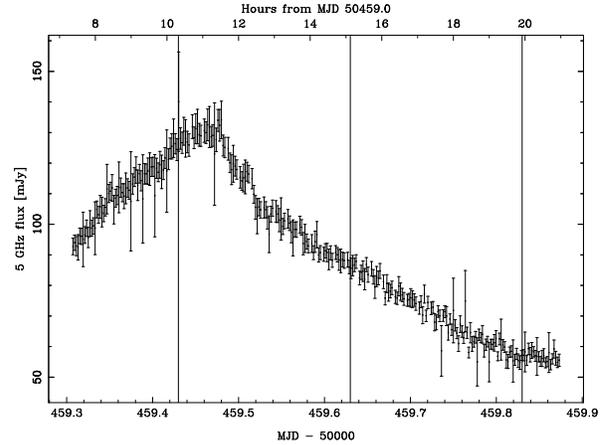}
\caption{Cyg X-3 photometry on 11 January 1997.  Time bins are 3.4 min.}
\label{photometry_6}
\end{figure}

For these observations the success of the MERLIN array is not in its
imaging capabilities, but its photometry.  Figs
\ref{photometry_1}-\ref{photometry_6} show photometry for all six C-band
epochs.  Note that in these figures the vertical scales vary and the
zero is not included.  Vertical lines in all these figures indicate
the time at which X-ray minimum occurs using the ephermeris by Matz et
al.\ (1996).
\begin{figure}
\epsfig{file=fig_08.eps, angle=-90, width=3.5in}
\caption{Cyg X-3 photometry on 28 December 1996.  Time bins are 2.0 min.}
\label{photometry_5}
\end{figure}
Data have been averaged into time bins of 2.0-4.0 minute duration to
reduce noise and all baselines have been used.

A number of features are visible in the plots.  Points to note in the
data are:
\begin{enumerate}
\item{In all but the first four hours of the MJD 50445 (28 December
1996) observation, the flux is variable on timescales shorter than a
couple of hours.  This causes problems in the mapping as discussed
above.}
\item{A number of flares occurred, however, the timescales involved are
not constant or correlated with flare strength or length of time in
quiescence preceding a flare.}
\end{enumerate}

\subsection{Power spectra}

\begin{figure}
\epsfig{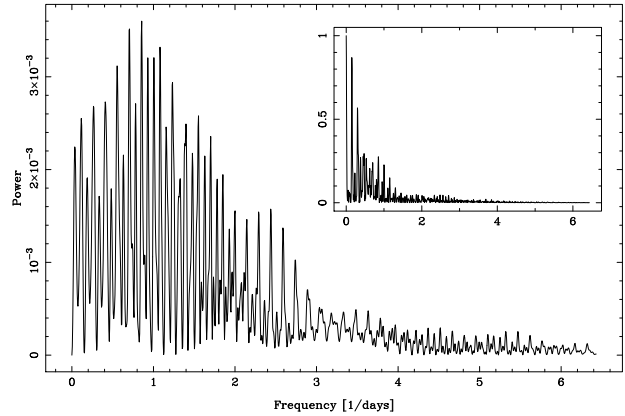}
\caption{A raw power spectrum from all six epoch photometric points.
The insert is the Fourier transform of the sampling function of the data.}
\label{Power_spectrum}
\end{figure}

To search for any periodic signals in the data, photometric points
from all six epochs were included and the Fourier spectrum was
determined.  This is shown in Fig.\ \ref{Power_spectrum}.
The main part of the figure is a direct Fourier transform of the
data.  The insert is the Fourier transform of the windowing function.

\begin{figure}
\epsfig{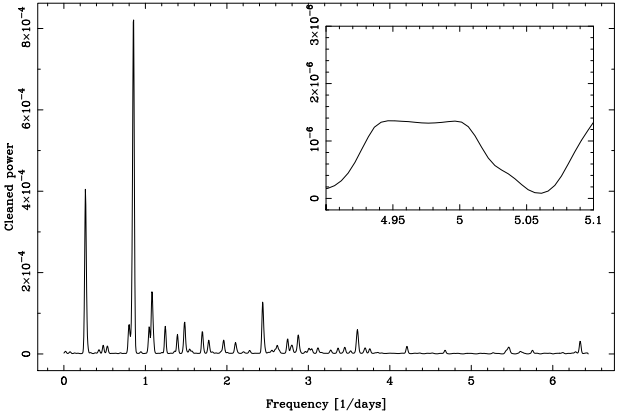}
\caption{A cleaned spectrum.  The insert shows the power at a period
of 4.8 hours.  There is no evidence for any periodic signal at this
binary period.}
\label{clean_spectrum}
\end{figure}

The main transform of the data is the time-domain equivalent of an
interferometric dirty map, and the transform of the windowing function
is the equivalent of a dirty beam.  In the same way as one can
re-create a clean map from a dirty map and dirty beam, the power
spectrum can be cleaned by an iterative subtraction of the window
function (Clark 1980; Roberts, Lehar \& Dreher 1987).  The result of
this procedure is shown in Fig.\ \ref{clean_spectrum}.

\begin{figure}
\epsfig{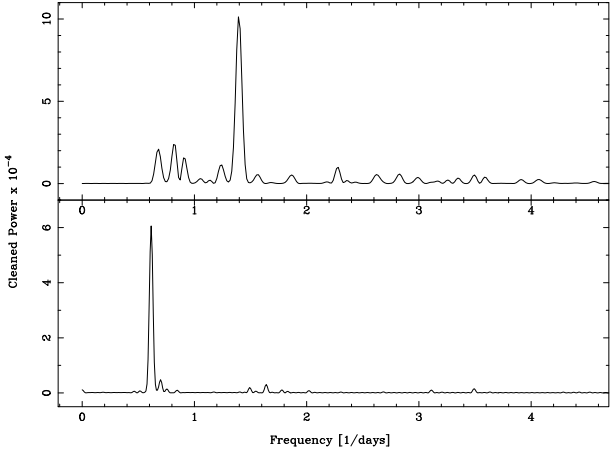}
\caption{Power spectra of the data cleaned in an identical way to
Fig.\ \ref{clean_spectrum}, but split into two halves.  The top
panel shows the first three epochs, and the bottom panel shows the
latter three epochs.  No correlation in the periods are found between
the two halves of the data.}
\label{clean_spectrum_split}
\end{figure}

To check whether the periods shown in Fig.\ \ref{clean_spectrum} are
spurious, the first three epochs were analysed separately from the
last three epochs.  The clean spectra are shown in Fig.\
\ref{clean_spectrum_split}.  As there is no correlation between the
two halves of the data sets, we can conclude that there are no stable
periodicities and any periodicity shown is due to spurious features or
quasi-periodic oscillations.

\section{Discussion}

\subsection{Brightness temperature of the flares}

A brightness temperature of a flare can be calculated associated with
the flux change from the source.  A black body sphere with radius $r$,
flux $S_{\nu}$, at a distance $D$, assuming a Rayleigh-Jeans tail,
would have a brightness temperature of
\begin{equation}
T_{\rm b} = \frac{c^{2}S_{\nu}D^{2}}{2 k_{\rm B}\nu^{2}r^{2}}
\end{equation}
where $c$ is the speed of light, $k_{\rm B}$ is Boltzmann's constant
and $\nu$ is the frequency.  Therefore a flare of duration $\Delta t$ minutes and flux change of $\Delta S$ mJy has a brightness temperature of 
\begin{equation}
T_{\rm b} = 9.66 \times 10^{10}\; \frac{\Delta S_{\rm mJy}
D^{2}_{\rm kpc}} {\nu_{\rm GHz}^{2} \Delta t_{\rm min}^{2}} \;\;{\rm K}
\end{equation}
where quantities are measured in their subscripted units.

For a typical flare: MJD 50419.56, the flux increase $\Delta S = 80$
mJy over $\Delta t = 46$ minutes at $\nu = 5$ GHz. If Cyg X-3 is at a
distance of 10 kpc, the brightness temperature is $T_{\rm b} \geq 1.5
\times 10^{10}$ K.  This value is typical for the majority of flares,
including the large flare at MJD 50432.  The largest brightness
temperature occurred at MJD 50438.6, if the change within 1
integration bin is believable.  For this flare the flux change is
$\Delta S = 10$ mJy over $\Delta t = 4.0$ min, and so the brightness
temperature is $T_{\rm b} \geq 2 \times 10^{11}$ K.

There is a maximum brightness temperature above which inverse Compton
losses become catastrophic.  This maximum brightness temperature can
be written for galactic sources as
\begin{equation}
T_{\rm max} \leq 1.6 \times 10^{12}\;
\left(c_{56}^{(\alpha)}\right)^{0.8} \left( \frac{\left(1 -
\alpha\right)} {\nu_{U}} \left[ \frac{\nu_{U}} {\nu} \right]^{\alpha}
\right)^{0.2}{\cal{D}}^{1.2},
\end{equation}
where $\nu_{U}$ is the upper frequency of the synchrotron emission in
GHz, $\cal{D}$ is the Doppler boosting factor of the emission and
$c_{56}^{(\alpha)}$ is the $\alpha$-dependent parts of the ratio of
Pacholczyk constants $c_{5}$ and $c_{6}$ where
\begin{equation}
c_{56}^{(\alpha)} = \frac{ 3\alpha + 5} {6\alpha^{2} + 19\alpha + 13}
\frac{\Gamma\left( \frac{3\alpha+1} {6} \right) \Gamma\left( \frac{3\alpha +
5} {6}\right )} { \Gamma\left( \frac{6\alpha + 5} {12} \right)
\Gamma\left( \frac{6\alpha+13} {12}\right)}.
\end{equation}
(Pacholczyk 1970; Hughes \& Miller 1991).  For a spectral index of
$\alpha = -0.1$, and $\nu_{U} = 350$ GHz (Ogley et al. in prep.), and a
frequency of $\nu = 5$ GHz used in this work, the expression for the
maximum brightness temperature evaluates to
\begin{equation}
T_{\rm max} \leq 4.8 \times 10^{11} {\cal{D}}^{1.2}\;\;\rm K.
\end{equation}
An estimation of the bulk Doppler factor of the jets can be inferred
from jet expansion.  An arcsec-scale measurement (Mart\'{\i} et al.\ 2000)
has shown that at that distance from the core, components are
expanding with proper motions of $\mu_a =10.9$ mas d$^{-1}$ and $\mu_r
= 9.8$ mas d$^{-1}$ for the approaching and receding components
respectively.  This allows an estimate of the jet bulk velocity and
angle to the line of sight to be measured, where
\begin{equation}
\beta\cos{\theta} = \frac{\mu_a - \mu_r}{\mu_a + \mu_r}.
\end{equation}
From these observations the authors find $\beta\cos{\theta} = 0.053
\pm 0.027$, from which they assume a jet velocity of $\beta = 0.60$
and an angle to the line of sight of $\theta = 85\degr$.  If these
values are adopted then a Doppler factor of ${\cal{D}} = 1.18$ can be
calculated.  If the jet speed is increased to $\beta = 0.89$ as
observed in VLBA observations (Mioduszewski et al. 2000), then by
using the same value of $\theta$, the Doppler factor increase to
${\cal{D}} = 2.02$.  Using this range of Doppler factors, the maximum
brightness temperature is therefore $T_{\rm max} \leq 6-11 \times
10^{11}$ K.

\subsection{Orbital motion}

The effect of the binary's orbital motion on the observed radio
emission from Cyg X-3 is found to be negligible, unlike Cyg X-1
(Brocksopp et al 1999).  Reasons why this could be are 
\begin{itemize}
\item[(i)]{The binary is face on (which is irrespective of the jet angle).
Throughout the orbit there is no difference in the amount of material
that the jets encounter which would scatter or absorb radiation.}
\item[(ii)]{Radio emission in Cyg X-3 is dominated by optically thin events
which occur from a region way outside the orbital diameter, and so any
4.8 h variation, in say particle production or blob ejection, is
smoothed out by the time we see it.  In Cyg X-1, the radio emission is
self-absorbed and is dominated by the high frequencies, this occurring
closer to the binary centre.}
\end{itemize}

To quantify this absence of radio emission, each run was Fourier
transformed into 100 frequencies with a separation of 0.1 cycles per
day.  The power spectra were then summed over all the data sets.  No
significant peak at or around 5 cycles per day ($\equiv$ 4.8 h period)
was found in either the individual spectra or in the sum.  A 2
$\sigma$ upper limit of 15 mJy can be set for the amplitude of any
sinusoidal variation in source flux for the average power spectrum.
Each daily run was also split into 5 overlapping data sets,
transformed and averaged to give an average power sepctrum per
run. These showed significantly higher power at low frequencies with a
power-law dependence with a typical slope of $-2.4 \pm 0.2$ over the
frequency range 4--40 cycles per day. The power law shows that no
particular timescale can be assigned to the data.

We have indicated on the figures \ref{photometry_1} --
\ref{photometry_6}, by vertical lines, the times of X-ray orbital
minimum as calculated from the ephemeris of Matz et al.\ (1996), and
one can see that any radio state is equally probable.

\subsection{Emission position}

From the small time-scale variations in the photometric data flares
can be identified as originating from a region typically 10 light
minutes, or 1.22 AU across.  However, it is highly unlikely that this
emission occurs close to the centre of the system where the jets are
accelerated, as both the jets and the compact object are buried deep
within an optically thick, dense stellar wind from the companion star,
see Ogley, Bell Burnell, Fender (2001).

The varying opacity to the radiation from the wind creates zones along
the jet where radiation at different frequencies dominates, as the
surrounding medium becomes optically thin.  Radii at which this occurs
can be calculated using a modified Wright \& Barlow (1975) wind model,
a function for the Gaunt factor appropriate for radio wavelengths
(Leitherer \& Robert 1991) and assuming spherical symmetry.  The
radius at which emission occurs then depends on only the wind
parameters such as composition, temperature and mass-loss rate, as
well as the frequency of the observed emission.  From the delay of
flares at 2.25, 8.3 and 15 GHz during 1991 to 1995, parameters for the
mean number of free electrons, $\gamma = 1$, the RMS ionic charge, $Z
= 1$, the mean atomic weight per nucleon, $\mu = 4$, the temperature,
$T = 20,000$ and the mass-loss rate, $\dot{M} = 1 \times 10^{-5}$
M$_{\odot}$ yr$^{-1}$ can be calculated (Waltman et al.\ 1996).  This
then simplifies the frequency-dependent radius to
\begin{equation}
R_{c} = 3.28 \times 10^{3}\;g^{1/3}\nu_{\rm GHz}^{-2/3}\;\;{\rm R_{\odot}},
\end{equation}
where
\begin{equation}
g = 9.77 + 1.27\;\log{\left(2.83 \times 10^{3}\;\nu_{\rm
GHz}^{-1}\right)}
\end{equation}
and $\nu_{\rm GHz}$ is the frequency measured in GHz.  For a frequency
of 5 GHz, the radius at which emission is observed is therefore $R_{c}
\geq$ 13 AU.  This creates significant layering in the emission from
the jets with the higher frequencies occurring closer to the centre
(at 350 GHz, emission would occur at only 0.5 AU; see Ogley et al.\ in
prep.).  We note that the Waltman et al mass loss rate assumes an
expansion velocity of only 0.3c; a higher velocity would give a larger
mass loss rate and larger values of $R_{c}$. Furthermore Ogley et al (2001)
find a significantly larger mass loss rate from infrared observations,
a rate independent of the expansion velocity. Finally we note that the
wind may not be spherically symmetric - see Ogley, Bell Burnell,
Fender (2001) and Fender, Hanson, Pooley (1999).

\section{Conclusions}

We have observed the X-ray binary Cygnus X-3 with the MERLIN
interferometer with the intention of mapping jet components in a
period of minor flare activity.  We observed for six epochs between 01 
December 1996 and 11 January 1997.  In all epochs the source showed
some degree of variability and consequently we were unable to map
the source without the creation of artifacts.

High time-resolution photometric observations show a number of
small-flux flares of around an hour in duration, with a large amount
of structure.  This is typical of a minor-flare period.  Power spectra 
of the data show no persistent periodicities in the data, and no
evidence of the orbital period of 4.8 hr.

A measurement of the brightness temperature for the flares show
typical values of $10^{9}$-$10^{10}$ K, with the largest values
occurring over 4 minutes and a brightness temperature of $2 \times
10^{11}$ K.  The flare emission is from a region of typically 10 light
minutes across, with a diameter of 1.22 AU.  However, to be visible at
a frequency of 5 GHz these plasmons would be situated at a distance of
13 AU from the core, assuming a spherical wind.

\section{Acknowledgments}

The Authors wish to thank various MERLIN staff for help with data and
analysis.  We are grateful for the help given by Tom Muxlow, Simon
Garrington and Peter Thomasson for scheduling the observations and
giving advice in the data reduction.  The authors also wish to thank
Duncan Law-Green for his assistance in attempting to map the data.
RNO wishes to thank Patrick McGrough and Robin Sanderson for their
hospitality during the writing of this paper.

Guy Pooley generously made available the Ryle Telescope data.

The Green Bank Interferometer is a facility of the National Science
Foundation operated by the NRAO in support of NASA High Energy
Astrophysics programme.  MERLIN is a National Facility operated by the
University of Manchester on behalf of PPARC.

\end{document}